\documentclass[british]{article}
\usepackage[T1]{fontenc}
\usepackage[latin9]{inputenc}
\usepackage{color}
\usepackage{babel}
\usepackage{url}
\usepackage{amsmath}
\usepackage{amssymb}
\usepackage{xargs}[2008/03/08]
\usepackage[unicode=true,
 bookmarks=true,bookmarksnumbered=false,bookmarksopen=false,
 breaklinks=false,pdfborder={0 0 0},backref=page,colorlinks=true]
 {hyperref}
\hypersetup{pdftitle={Kernel Quantum Probabilities},
 pdfauthor={B. Piwowarski},
 pdfkeywords={quantum probabilities, kernel}}

\makeatletter
\usepackage{float}
\usepackage[noend]{algorithmic}
\floatstyle{ruled}
\newfloat{algorithm}{tbp}{loa}
\floatname{algorithm}{Algorithm}

 \newcommand*{\system}[1]{{\ttfamily #1}}

\@ifundefined{date}{}{\date{}}
\usepackage{fullpage}
\usepackage{mathrsfs}
\usepackage{fancyvrb}
\usepackage{color}

\makeatother

\begin{document}

\title{The Kernel Quantum Probabilities (KQP) Library}

\author{B. Piwowarski\\
\texttt{\footnotesize benjamin@bpiwowar.net}}
\maketitle
\begin{abstract}
In this document, we show how the different quantities necessary to
compute kernel quantum probabilities can be computed. This document
form the basis of the implementation of the Kernel Quantum Probability
(KQP) open source project%
\footnote{\url{http://kqp.bpiwowar.net/}%
}.
\end{abstract}
\global\long\def\units{\mathcal{U}}
\global\long\def\normv{\left\Vert \mathcal{V}\right\Vert }
\global\long\def\dproj#1#2{#1\vartriangleright#2}
\global\long\def\Projector{\mathcal{P}}

\global\long\def\bra#1{#1^{\dagger}}
\global\long\def\ket#1{#1}
\global\long\def\braAket#1#2#3{\left\langle #1\middle|#2\middle|#3\right\rangle }

\global\long\def\braket#1#2{\left\langle #1\middle|#2\right\rangle }
\global\long\def\pr{\mbox{Pr}}
\global\long\def\event#1{\mathbf{#1}}
\global\long\def\kbasis#1{\ket{#1}}
\global\long\def\bbasis#1{\bra{#1}}

\global\long\def\norm#1{\left\Vert #1\right\Vert }
\global\long\def\card#1{\left|#1\right|}

\global\long\def\argmax{\mathop{\arg\max}}

\global\long\def\argmin{\mathop{\arg\min}}

\global\long\def\st{\mbox{ s.t. }}

\global\long\def\vproj#1{\ket{#1}\bra{#1}}
\global\long\def\kt{\middle|}

\global\long\def\trtimes#1{\mathcal{P}\left(#1\right)}

\global\long\def\Hilbert{\mathcal{H}}
\newcommandx\FMatrix[1][usedefault, addprefix=\global, 1=]{\mathcal{M}_{\Hilbert}^{#1}}
\global\long\def\Matrix{\mathcal{M}}

\global\long\def\Id{Id}
\global\long\def\tr{\mbox{tr}}

\global\long\def\dproj#1#2{#1\triangleright#2}

\global\long\def\X{\mathcal{X}}
\global\long\def\U{\mathcal{U}}
\global\long\def\V{\mathcal{V}}
\global\long\def\W{\mathcal{W}}
\global\long\def\Y{\mathcal{Y}}

\global\long\def\fvectors{\mathscr{U}}
\global\long\def\foperator{\mathfrak{U}}
$ $

\global\long\def\XP{\overline{\X}}
\global\long\def\UP{\overline{\U}}
\global\long\def\repr{\psi}

\global\long\def\nth{{\tiny\mathrm{th}}}

\global\long\def\oequal#1{\stackrel{\mathbf{o}[#1]}{=}}
\global\long\def\second{\prime\prime}

\input{pygment}

\section{Introduction}

Quantum Probabilities correspond to one of the generalisation of standard
probabilities. It is founded on the mathematical theory underlying
Quantum Physics. This framework was developed in the 1930s by von
Neumann and Dirac. It was recently further developed and generalised
by the so-called ``sequential effect algebra'' \cite{Gudder2007Quantum-Probability}.
The Kernel Quantum Probability library (KQP) aims to provide tools
to effectively compute ``quantum probabilities'', that is to compute
a representation of densities, events and to update the densities
when events are observed (conditionalisation). It also provides access
to generalisation of standard probabilistic measure like entropy and
divergence \cite{Gudder2007Quantum-Probability}.

Computing quantum probabilities related quantities relies on linear
algebra, and more precisely on the definition of an inner product
in a Hilbert space, since this defines the probability of transition
(when measuring) between possible system states. 

In the machine learning community, a standard ``trick'' is to use
a \emph{kernel} to define the inner product \cite{Scholkopf2002Learning-with}.
That is, states can be represented in an arbitrary feature space $\mathcal{F}$
for which there exists a mapping $\Phi$ such that $\Phi(x)\cdot\Phi\left(y\right)$
is valid inner product. We call $k(x,y)=\Phi(x)\cdot\Phi(y)$ the
kernel, which can be computed without explicitly computing $\Phi(x)$,
thus allowing to work in high or infinite spaces. 

This documents describe how to compute quantum probabilities related
quantities relying only on the inner product definition given by the
kernel. The organisation of this document is as follows:
\begin{enumerate}
\item In Section~\ref{sec:kqpr}, we describe how to compute probabilities
and how to update the probabilities given a subspace (or its orthogonal),
both from a theoretical point of view and implementation point of
view.
\item In Section~\ref{sec:kevd}, we describe how to compute an approximation
of \emph{quantum} densities or events.
\item In Section~\ref{sec:example}, we give an example of code using KQP
in C++.
\end{enumerate}

\section*{Notations}

We suppose that we work within a complex Hilbert space, that is that
the field is $\mathbb{C}$ unless otherwise specified. The set of
complex matrices of dimensions $n$ by $p$ is denoted $\Matrix_{n\times p}$.

In order to deal with kernels, following the literature, data points
in the original space will be called pre-images since they are used
to build a basis of the subspace containing the quantum density or
event (see Section \ref{sec:kqpr}). 

In order to use common linear algebra notations, we consider a list
of pre-images as a linear map, and use uppercase calligraphic letters
to denote a list of pre-images: A list $\X$ of $n$ pre-images is
denoted

\[
\X\in\FMatrix[(n)]
\]
An arbitrary list belongs to $\FMatrix=\bigcup_{n}\FMatrix[(n)]$.
A linear combination of pre-images is simply denoted $\X A$ where
$A\in\Matrix_{n\times p}$.

We define the adjoint operator in a natural way, i.e. it maps a list
of pre-images into 
\[
\X^{\dagger}\in\cup_{p}\mathcal{L}\left(\FMatrix[(p)];\mathcal{M}_{n\times p}\right)
\]
We denote $k$ the kernel, i.e. we denote 
\[
k(\X,\U)=\X^{\dagger}\U\in\mathcal{M}_{n\times p}
\]
with $\X\in\FMatrix[(n)]$ and $\U\in\FMatrix[(p)]$. 

Finally, we use the symbol $\Projector$ to denote the composition
of a linear operator with its transpose, i.e. 
\[
\Projector\left(A\right)=AA^{\dagger}
\]

\section{Computing probabilities}

\label{sec:kqpr} Readers are referred to \cite{Gudder2007Quantum-Probability,Pitowsky:1989kr}
for a discussion and presentation of what are quantum probabilities.
Shortly, they can be defined by:
\begin{description}
\item [{A~quantum~density}] is a positive semi-definite self-adjoint
linear operator $\rho$ of trace 1;
\item [{An~Observable}] is a projector and corresponds to a yes-no measurement,
i.e. to a quantum ``event'';
\item [{An~effect}] is an operator $A$ such that $0\le\left\langle Ax,x\right\rangle \le1$.
Note that an observable is an effect, but the reverse is not true.
Effects can be considered as ``fuzzy'' or ``imprecise'' observables.
\end{description}
In this section, we present the formulas corresponding to the various
quantities of interest (probability, conditionalisation, divergence)
and how we can compute them within KQP. This section is based on the
work of Gudder \cite{Gudder2007Quantum-Probability} (effects) and
\cite{Umegaki:1962ee} (divergence and entropy).

In this section, we suppose we have a density $\rho$ and an effect
$E\in\mathcal{E}\left(\Hilbert\right)$ that can be decomposed as:
\[
\rho=\Projector\left(\X_{\rho}Y_{\rho}\Sigma_{\rho}\right)\mbox{ and }E=\Projector\left(\X_{E}Y_{E}\Sigma_{E}\right)
\]
where $\X_{\bullet}$ belong to $\FMatrix$ ,$Y_{\bullet}$ and $Z_{\bullet}$
to $\Matrix$.  

For some operations, we need the decomposition to be in an orthonormal
form, i.e. that 
\begin{equation}
Y^{\dagger}\X^{\dagger}\X Y=\Id\label{eq:ortho}
\end{equation}

When using the orthonormality hypothesis, we use the symbol $\mathbf{o}[\ldots]$
over the equality. For example,

\[
Y_{\rho}^{\dagger}\X_{\rho}^{\dagger}\X_{\rho}Y_{\rho}\oequal{\rho}\Id
\]

Finally, for densities we use the proportionality to denote that it
should be normalised, i.e. $\rho\propto\rho_{u}$ means that 
\[
\rho=\frac{\rho_{u}}{\mbox{tr}\left(\rho_{u}\right)}
\]
Note that it is straightforward to compute the normalisation factor,
since, using the cyclic re-ordering property of trace operators:
\[
\tr\left(\rho\right)=\tr\left(\X_{\rho}Y_{\rho}\Sigma_{\rho}^{2}Y_{\rho}^{\dagger}\X_{\rho}^{\dagger}\right)=\tr\left(\Sigma_{\rho}^{2}Y_{\rho}^{\dagger}\X_{\rho}^{\dagger}\X_{\rho}Y_{\rho}\right)
\]
If the decomposition is orthonormal (Eq. \ref{eq:ortho}), we have
$\tr\left(\rho\right)=\tr\left(\Sigma_{\rho}^{2}\right)=\left\Vert \Sigma_{\rho}\right\Vert ^{2}$.

For a matrix $A$, we denote $A_{\bullet j}$ its $j^{\nth}$ column,$A_{i\bullet}$
its $i^{\nth}$ row. If a matrix $A$ as an subscript $\rho$, we
use a semicolon to separate the subscript from the column/row indices,
as for example in $A_{\rho;ij}$.

\subsection{Computing probabilities}

The probability of an effect $E$ is defined as

\[
\pr_{\rho}\left(E\right)=\mbox{tr}\left(\rho E\right)
\]
We can compute the probability of an effect $E$ using the re-ordering
property of the trace operator

\[
\pr_{\rho}\left(E\right)=\mbox{tr}\left(\rho E\right)=\left\Vert \Sigma_{E}Y_{E}^{\dagger}k\left(\X_{E},\X_{\rho}\right)Y_{\rho}\Sigma_{\rho}\right\Vert ^{2}
\]

\subsection{Entropy}

The entropy of a density $\rho$ (with an orthonormal decomposition)
can be written \cite{Umegaki:1962ee}:

\begin{align*}
\tr(\rho\log(\rho))) & \oequal{\rho}\tr\left(\left(Y_{\rho}^{\dagger}\X_{\rho}^{\dagger}\X_{\rho}Y_{\rho}\right)\Sigma_{\rho}\left(Y_{\rho}^{\dagger}\X_{\rho}^{\dagger}\X_{\rho}Y_{\rho}\right)\log\left(\Sigma_{\rho}^{2}\right)\right)\\
 & =\tr\left(\Sigma_{\rho}^{2}\log\left(\Sigma_{\rho}^{2}\right)\right)\\
 & =\sum_{i}2\Sigma_{\rho;ii}^{2}\log\left(\Sigma_{\rho;ii}\right)
\end{align*}

\subsection{Divergence}

Umegaki \cite{Umegaki:1962ee} proved that the equivalent of the Kullback-Leilbler
divergence between two densities $\rho$ and $\tau$ can be computed
as: 
\[
J(\rho||\tau)=\tr(\rho\log(\rho)-\rho\log(\tau)))
\]

The first part corresponds to the entropy, and the second part can
be computed as follows. In order to deal with infinities, in practice
we want to compute the divergence using $\tau^{\prime}=\left(1-\epsilon\right)\tau+\epsilon\alpha\Id$
where $\alpha\Id$ is a blank noise, i.e. $\tr\left(\alpha\Id\right)=1$.
In case of infinities, $\alpha$ can be set to a small value. We have:

\begin{align*}
\tr\left(\rho\log\tau^{\prime}\right) & =\tr\left(\rho\log\left(\left(1-\epsilon\right)\tau+\epsilon\alpha\Id\right)\right)\\
 & \oequal{\tau}\tr\left(\rho\left[\X_{\tau}Y_{\tau}\log\left(\left(1-\epsilon\right)\Sigma_{\tau}^{2}+\epsilon\alpha\Id\right)Y_{\tau}^{\dagger}\X_{\tau}^{\dagger}+\log\left(\epsilon\alpha\right)\left(\Id-\X_{\tau}Y_{\tau}Y_{\tau}^{\dagger}\X_{\tau}^{\dagger}\right)\right]\right)\\
 & =\tr\left(\Sigma_{\rho}Y_{\rho}^{\dagger}\X_{\rho}^{\dagger}\X_{\tau}Y_{\tau}\log\left(\left(1-\epsilon\right)\Sigma_{\tau}^{2}+\epsilon\alpha\Id\right)Y_{\tau}^{\dagger}\X_{\tau}^{\dagger}\X_{\rho}Y_{\rho}\Sigma_{\rho}\right)+\log\left(\epsilon\alpha\right)\left(\tr\left(\rho\right)-\tr\left(\rho\X_{\tau}Y_{\tau}Y_{\tau}^{\dagger}\X_{\tau}^{\dagger}\right)\right)\\
 & =-\left\Vert \Sigma_{\rho}Y_{\rho}^{\dagger}\X_{\rho}^{\dagger}\X_{\tau}Y_{\tau}\left(\log\left(\left(1-\epsilon\right)\Sigma_{\tau}^{2}+\epsilon\alpha\Id\right)\right)^{1/2}\right\Vert ^{2}+\log\left(\epsilon\alpha\right)\left(1-\left\Vert \Sigma_{\rho}Y_{\rho}^{\dagger}\X_{\rho}^{\dagger}\X_{\tau}Y_{\tau}\right\Vert ^{2}\right)
\end{align*}

\subsection{Conditionalisation}

We first give the formulas to compute the conditional quantum density
when observing an effect $E$, and then when observing its orthogonal
$E^{\bot}$

\subsubsection{Projecting on the effect $E$}

If we observe the event $E$, the density $\rho$ conditioned upon
$E$, denoted $\rho\triangleright E$, is given by: \textbf{
\[
\rho\triangleright E=\frac{E^{1/2}\rho E^{1/2}}{\tr\left(\rho E\right)}
\]
}

We can focus on the numerator since we only have to normalise the
resulting density afterwards. We have 
\[
\dproj{\rho}E\propto E^{1/2}\rho E^{1/2}
\]

We can distinguish two cases:
\begin{enumerate}
\item $E$ is an observable: since $E=E^{1/2}$, we have 
\[
\dproj{\rho}E=\Projector\left[\X_{E}\left(Y_{E}\Sigma_{E}^{2}Y_{E}^{\dagger}k\left(\X_{E},\X_{\rho}\right)Y_{\rho}\right)\Sigma_{\rho}\right]
\]

\item $E$ is a ``strict'' effect: In this case, we require an orthonormal
decomposition for E, and we can compute the projection as:
\end{enumerate}
\[
\dproj{\rho}E\oequal E\Projector\left[\X_{E}\left(Y_{E}\Sigma_{E}Y_{E}^{\dagger}k\left(\X_{E},\X_{\rho}\right)Y_{\rho}\right)\Sigma_{\rho}\right]
\]

In both cases, the resulting density is not in an orthonormal form.

\subsubsection{Projecting on the orthogonal $E^{\bot}$}

If we observe the orthogonal of event $E$, we can update our knowledge
on $\rho$, denoted $\rho\triangleright E^{\bot}$, as:
\[
\dproj{\text{\ensuremath{\rho}}}{E^{\bot}}=\frac{\left(\Id-E\right)^{1/2}\rho\left(\Id-E\right)^{1/2}}{1-\tr\left(\rho E\right)}
\]

When $E$ is in an orthonormal form, we can use the fact that $\Id-Y_{E}\X_{E}\X_{E}^{\dagger}Y_{E}^{\dagger}$
is the projector on the space orthogonal to the space spanned by the
vectors of $E$. Thus,

\begin{eqnarray}
\left(\Id-E\right)^{1/2} & = & \X_{E}Y_{E}\left(\Id-\Sigma^{2}\right)^{1/2}Y_{E}^{\dagger}\X_{E}^{\dagger}+\left(\Id-\X_{E}Y_{E}Y_{E}^{\dagger}\X_{E}^{\dagger}\right)\nonumber \\
 & = & \Id-\X_{E}Y_{E}\left[\Id-\left(\Id-\Sigma_{E}^{2}\right)^{1/2}\right]Y_{E}^{\dagger}\X_{E}^{\dagger}\label{eq:effect-orthogonal}
\end{eqnarray}

Using the above, we can write:

\begin{align*}
\dproj{\rho}{E^{\bot}} & \propto\trtimes{\left(\Id-E\right)^{1/2}\X_{\rho}Y_{\rho}\Sigma_{\rho}}\\
 & \propto\Projector\left[\left(\begin{array}{cc}
\X_{\rho} & \X_{E}\end{array}\right)\left(\begin{array}{c}
\Id\\
-Y_{E}\left[\Id-\left(\Id-\Sigma_{E}^{2}\right)^{1/2}\right]Y_{E}^{\dagger}k\left(\X_{E},\X_{\text{\ensuremath{\rho}}}\right)
\end{array}\right)Y_{\rho}\Sigma_{\rho}\right]
\end{align*}

We readily verify that when $\Sigma_{E}=\Id$ it gives the right formula
$\rho-\X YY^{\dagger}X^{\dagger}\rho$. We can use in those cases
a direct EVD approach (section \ref{sub:direct-EVD}) to obtain a
simplified form.

\section{Approximating operators}

In this section, we describe the techniques used to computed low-rank
approximations of linear operators in the feature space. In particular,
we are interested in methods where the operator can be decomposed
as: 
\begin{equation}
\sum_{i}\alpha_{i}\U_{i}A_{i}A_{i}^{\dagger}\U_{i}^{\dagger}\approx\X Y\Sigma Y^{\dagger}\X^{\dagger}\label{eq:approx}
\end{equation}
 were $\X Y$ is (or might be) orthonormal, i.e. $Y^{\dagger}\X^{\text{\ensuremath{\dagger}}}\X^{\dagger}Y$
is the identity.

In the following, we describe:
\begin{itemize}
\item In Section \ref{sub:direct-EVD}, how to get an EVD decomposition
of any linear operator of the form $\foperator=\X AA^{\dagger}\X^{\dagger}$.
This is useful in order to e.g. lower the rank and is needed or before
removing feature space vectors from $\X$.
\item In Section \ref{sec:kevd}, we show how to update the EVD of a linear
operator $\foperator$ with a low rank operator $\alpha_{i}\U_{i}A_{i}A_{i}^{\dagger}\U_{i}^{\dagger}$.
\item In Section \ref{sec:reduced-set}, we show how to remove feature vectors
from $\X$ when we have an EVD $\foperator=\X A\Sigma A^{\dagger}\X^{\dagger}$.
We use two techniques: 

\begin{itemize}
\item Null space method (Section \ref{sub:null-space-method})
\item Quadratic optimisation to find the subset of pre-images that minimise
the reconstruction error (Section \ref{sec:reduced-set}).
\end{itemize}
\end{itemize}

\subsection{Direct EVD}

\label{sub:direct-EVD}

In this section, we discuss how to get the orthonormal form of an
operator written as $\foperator=\X ASA^{\dagger}\X^{\dagger}$ where
$S$ is a diagonal matrix. We first describe the case where $S$ is
positive semi-definite, before tackling the general case. 

This type of approach is useful in several cases, and the builder
\system{AccumulatorKernelEVD} in KQP relies on this decomposition,
since it represents Eq. (\ref{eq:approx}) as
\[
\left(\begin{array}{ccc}
\U_{1} & \cdots & \U_{n}\end{array}\right)\left(\begin{array}{ccc}
A_{1}\\
 & \ddots\\
 &  & A_{n}
\end{array}\right)\left(\begin{array}{ccc}
\Sigma_{1}\\
 & \ddots\\
 &  & \Sigma_{n}
\end{array}\right)\left(\begin{array}{ccc}
A_{1}\\
 & \ddots\\
 &  & A_{n}
\end{array}\right)^{\dagger}\left(\begin{array}{ccc}
\U_{1} & \cdots & \U_{n}\end{array}\right)^{\dagger}
\]
where $\Sigma_{i}=\mbox{diag}\left(\alpha_{i},\ldots,\alpha_{i}\right)$.

\subsubsection{Semi-positive definite case}

Suppose we have $\foperator=\X AA^{\dagger}\X^{\dagger}$ and we wish
to transform it to an orthonormal form. To achieve this, we have to
compute a thin EVD 
\[
EDE^{\dagger}=A^{\dagger}\X^{\dagger}\X A
\]
It is then straightforward to obtain the desired form by posing $Y=AED^{-1/2}$
and $\Sigma=D^{1/2}D^{1/2}$ 
\[
\X Y\Sigma Y^{\dagger}\X^{\dagger}=\X AEE^{\dagger}A^{\dagger}\X^{\text{\ensuremath{\dagger}}}=\X ASA^{\dagger}\X^{\dagger}
\]
where the last equality can be shown has follows. Any vector $y\in\Hilbert$
can be written $\X AP+\V Q$ where $\V^{\dagger}\X A=0$. Then, 
\begin{align*}
\X AEE^{\dagger}A^{\dagger}\X^{\text{\ensuremath{\dagger}}}y & =\X A\underbrace{EE^{\dagger}\underbrace{A^{\dagger}\X^{\text{\ensuremath{\dagger}}}\X A}_{EDE^{\dagger}}}_{A^{\dagger}\X^{\dagger}\X A}P+\underbrace{\X AEE^{\dagger}A^{\dagger}\X^{\dagger}\V Q}_{0}\\
 & =\X AA^{\dagger}\X^{\dagger}\X AP+\underbrace{\X AA^{\dagger}\X^{\dagger}\V Q}_{0}\\
 & =\X AA^{\dagger}\X^{\dagger}y
\end{align*}

We also can show easily that $\X Y$ is an orthonormal matrix 
\[
Y^{\dagger}\X^{\dagger}\X Y=D^{-1/2}E^{\dagger}\left(A^{\dagger}\X^{\dagger}\X A\right)ED^{-1/2}=\Id
\]

It is then possible to remove some pre-images using techniques from
Section \ref{sec:reduced-set}.

\subsubsection{General case}

In the general case, we have $\foperator=\X ASA^{\dagger}\X^{\dagger}$
which can be rewritten $\foperator=\X AS^{1/2}S^{1/2}A^{\dagger}\X^{\dagger}$.
That is, unless we use a real field and $S$ is not semidefinite positive.
In that case, we can still write
\[
\foperator=\X BB^{\dagger}\X^{\dagger}-2\X CC^{\dagger}\X^{\dagger}
\]
 where 
\begin{align*}
B & =A\left(S_{+}+S_{-}\right)^{1/2}\\
C & =AS_{-}^{1/2}
\end{align*}
where $S_{\pm}$ is the $S$ matrix where negative (resp. positive)
values are set to 0. We then use the approach above to compute an
orthonormal decomposition of $\X BB^{\dagger}\X^{\dagger}$, and then,
using the fact that the space defined by $\X YY\X^{\dagger}$ contains
$\X C$, 
\begin{align*}
\foperator & =\X Y\Sigma Y^{\dagger}\X^{\dagger}-2\X CC^{\dagger}\X^{\dagger}\\
 & =\X Y\left[\Sigma-2ZZ^{\dagger}\right]Y^{\dagger}\X^{\dagger}
\end{align*}
with $Z=Y^{\dagger}\X^{\dagger}\X C$. We then have to compute another
EVD for $\Sigma-2ZZ^{\dagger}$, which will give the final form of
$ $$\foperator$.

\subsection{Low-rank update of operators}

\label{sec:kevd}

The problem is to compute a low rank approximation of 
\[
\foperator=\sum_{i}\alpha_{i}\U_{i}A_{i}A_{i}^{\dagger}\U_{i}^{\dagger}
\]
where $\U_{i}\in\FMatrix$. 

In the following, we consider just one update and we drop the $i$
for more clarity. We further assume that we have a current approximation
decomposition expressed as 
\[
\foperator=\X YZ\Sigma Z^{\dagger}Y^{\dagger}\X
\]
where
\begin{itemize}
\item $\X\in\FMatrix[(n)]$ and $Y$ is a $n\times r$ matrix such that
$\X Y$ is orthonormal;
\item $Z$ is a $r\times r$ unitary matrix. This matrix is used in order
to avoid updating the potentially larger matrix $Y$ when the list
of pre-images remain the same;
\item $\Sigma$ is a diagonal matrix of rank $r$
\end{itemize}
In order to be able to process incrementally the set of vectors $\units$,
we wish to compute at each step a rank one update of $U$
\[
\widehat{\text{\ensuremath{\foperator}}}=\foperator+\alpha\U AA^{\dagger}\U^{\dagger}\approx\widetilde{\foperator}=\X^{\prime}Y^{\prime}Z^{\prime}\Sigma\left(\X^{\prime}Y^{\prime}Z^{\prime}\right)^{\dagger}
\]
This problem is related to~\cite{Chin2006Incremental-Kernel} that
deals with incremental Kernel SVD, and we follow mainly the same approach.
We use the following constraints:
\begin{enumerate}
\item Keep the (relative) error $\epsilon=\left\Vert \widehat{\text{\ensuremath{\foperator}}}-\widetilde{\foperator}\right\Vert /\left\Vert \text{\ensuremath{\foperator}}\right\Vert $
below a limit $\eta$ (if possible, see below);
\item Keep the rank $r$ below the limit $r_{\max}$ ;
\item Keep the number of pre-images below a number $cr$ where $c\ge1$.
\end{enumerate}

\subsubsection{Pre-computations}

\label{sub:rank-n-EVD}

We can write $\U$ as the direct sum

\begin{equation}
\U A=\underbrace{\left(\Id-\X YY^{\dagger}\X^{\dagger}\right)\U A}_{\V}+\X Y\underbrace{Y^{\dagger}\X^{\dagger}\U A}_{W}\label{eq:def:v}
\end{equation}
The operator $W$ can be computed explicitly as:
\begin{equation}
W=Y^{\dagger}\X^{\dagger}\U A=Y^{\dagger}k\left(\X,\U\right)A\label{eq:def:W}
\end{equation}

\subparagraph{}

\subparagraph{General case}

We can compute $\V^{\dagger}\V$ as
\[
\V^{\dagger}\V=A^{\dagger}k\left(\U,U\right)A-W^{\dagger}W
\]
which can in turn be used to compute%
\footnote{Note that we could use a Cholesky decomposition $k\left(\U,\U\right)=LL^{\dagger}$
followed by a generalised SVD on $L^{\dagger}A$ and $W$ to find
the EVD of $\V\V^{\dagger}$.%
} the (full) EVD of $\V\V^{\dagger}$.

\subparagraph{Special case $\U=\X$}

When $\U A$ is a linear combination of kernel vectors. In this case,
we have $\V=0$ and 
\[
W=Y^{\dagger}\X^{\dagger}\X A=Y^{\dagger}k\left(\X,\X\right)A
\]

\paragraph{Updating the operator}

Let us express $\U A$ as the direct sum of its projection onto the
subspace spanned by $\X YY^{\dagger}\X^{\dagger}$ and its orthogonal.
Since by definition $W=\left(\X Y\right)^{\dagger}\U A$, we can write
$\U A$ as:
\[
\U AA^{\dagger}\U^{\dagger}=\Projector\left(\X YY^{\dagger}\X^{\dagger}\U A+\V\right)=\Projector\left(\left(\begin{array}{cc}
\X & \V\end{array}\right)\left(\begin{array}{cc}
Y & 0\\
0 & Q
\end{array}\right)\left(\begin{array}{cc}
WQD^{1/2} & WQ_{0}\\
D^{1/2} & \mathbf{0}
\end{array}\right)\right)
\]
where $Q$, $Q_{0}$ and $D$ such that $\left(\begin{array}{cc}
Q & Q_{0}\end{array}\right)$ is unitary and 
\begin{equation}
\V QDQ^{\dagger}\V^{\dagger}=\V\V^{\dagger}\label{eq:constraint-q-1}
\end{equation}
We can write

\[
\foperator=\Projector\left(\begin{array}{cc}
\X YZ\Sigma^{1/2} & 0\end{array}\right)=\Projector\left[\left(\begin{array}{cc}
\X & \V\end{array}\right)\left(\begin{array}{cc}
Y & 0\\
0 & Q
\end{array}\right)\left(\begin{array}{cc}
Z & 0\\
0 & 0
\end{array}\right)\left(\begin{array}{cc}
\Sigma^{1/2} & 0\\
0 & 0
\end{array}\right)\right]
\]
and hence:

\begin{multline}
\foperator+\alpha\U AA^{\dagger}\U^{\dagger}=\left(\begin{array}{cc}
\X & \V\end{array}\right)\left(\begin{array}{cc}
Y & 0\\
0 & Q
\end{array}\right)\left[\left(\begin{array}{cc}
Z & 0\\
0 & 0
\end{array}\right)\left(\begin{array}{cc}
\Sigma & 0\\
0 & 0
\end{array}\right)\left(\begin{array}{cc}
Z & 0\\
0 & 0
\end{array}\right)^{\dagger}+\right.\\
\left.\alpha\left(\begin{array}{cc}
WQD^{1/2} & WQ_{0}\\
D^{1/2} & \mathbf{0}
\end{array}\right)\left(\begin{array}{cc}
WQD^{1/2} & WQ_{0}\\
D^{1/2} & \mathbf{0}
\end{array}\right)^{\dagger}\right]\left(\begin{array}{cc}
Y & 0\\
0 & Q
\end{array}\right)^{\dagger}\left(\begin{array}{cc}
\X & \mathcal{V}\end{array}\right)^{\dagger}\label{eq:U1}
\end{multline}

\paragraph{Computing $Q$}

Since $\left(\begin{array}{cc}
X & \V\end{array}\right)\left(\begin{array}{cc}
Y & 0\\
0 & Q
\end{array}\right)$ should be an orthonormal matrix, we should have:

\begin{eqnarray*}
\left(\left(\begin{array}{cc}
X & \V\end{array}\right)\left(\begin{array}{cc}
Y & 0\\
0 & Q
\end{array}\right)\right)^{\dagger}\left(\begin{array}{cc}
X & \V\end{array}\right)\left(\begin{array}{cc}
Y & 0\\
0 & Q
\end{array}\right) & = & \left(\begin{array}{c}
\left(\X Y\right)^{\dagger}\\
Q^{\dagger}\V^{\dagger}
\end{array}\right)\left(\begin{array}{cc}
\X Y & \V Q\end{array}\right)\\
 & = & \left(\begin{array}{cc}
\Id & 0\\
0 & Q^{\dagger}\V^{\dagger}\V Q
\end{array}\right)
\end{eqnarray*}
where we used the fact that $\left(\X Y\right)^{\dagger}\V=0$ from
Eq.~\eqref{eq:def:v}. Hence, orthonormality is equivalent to
\begin{align}
Q^{\dagger}\V^{\dagger}\V Q & =\Id\label{eq:constraint-q-2}
\end{align}

Using the result of Section \ref{sub:direct-EVD}, we can compute
the thin EVD $CDC^{\dagger}$ of $k\left(\V,\V\right)$ and pose $Q=CD^{-1/2}$
which will verify both Eqs. (\ref{eq:constraint-q-1}) and (\ref{eq:constraint-q-2})
. $Q_{0}$ corresponds to the basis of the null space obtained using
the same decomposition (note that if $\V=0$, $Q_{0}$ is the identity).

\paragraph{Updating}

We can use standard rank-one update techniques to update the decomposition;
since $Z$ is unitary, we can write 

\begin{align*}
\lefteqn{\left(\begin{array}{cc}
Z & 0\\
0 & \Id
\end{array}\right)\left(\begin{array}{cc}
\Sigma & 0\\
0 & 0
\end{array}\right)\left(\begin{array}{cc}
Z & 0\\
0 & \Id
\end{array}\right)^{\dagger}+\alpha\left(\begin{array}{cc}
WQD^{1/2} & WQ_{0}\\
D^{1/2} & \mathbf{0}
\end{array}\right)\left(\begin{array}{cc}
WQD^{1/2} & WQ_{0}\\
D^{1/2} & \mathbf{0}
\end{array}\right)^{\dagger}}\\
= & \left(\begin{array}{cc}
Z & 0\\
0 & \Id
\end{array}\right)\left(\left(\begin{array}{cc}
\Sigma & 0\\
0 & 0
\end{array}\right)+\alpha\left(\begin{array}{cc}
Z^{\dagger}WQD^{1/2} & WQ_{0}\\
D^{1/2} & \mathbf{0}
\end{array}\right)\left(\begin{array}{cc}
Z^{\dagger}WQD^{1/2} & WQ_{0}\\
D^{1/2} & \mathbf{0}
\end{array}\right)^{\dagger}\right)\left(\begin{array}{cc}
Z & 0\\
0 & \Id
\end{array}\right)^{\dagger}
\end{align*}
which is a rank $p$ update of a diagonal matrix. Note that, as we
cannot really compute $\V$, we have to get back to an expression
where $\U$ appears in the first matrix: 
\begin{eqnarray*}
\left(\begin{array}{cc}
\X & \V\end{array}\right)\left(\begin{array}{cc}
Y & 0\\
0 & Q
\end{array}\right) & = & \left(\begin{array}{cc}
\X & \U A-\X YW\end{array}\right)\left(\begin{array}{cc}
Y & 0\\
0 & Q
\end{array}\right)\\
 & = & \left(\begin{array}{cc}
X & \U\end{array}\right)\left(\begin{array}{cc}
Y & -YWQ\\
0 & AQ
\end{array}\right)
\end{eqnarray*}

\subsection{Reducing the pre-image set}

In this section, we describe the techniques used to reduce the number
of pre-images. When a linear combination of pre-images is possible,
it is better to use the direct EVD approach described in Section \ref{sub:direct-EVD}.

\subsubsection{Null space method}

\label{sub:null-space-method}

Suppose we have an operator $\foperator$ defined as
\[
\foperator=\X Y\Sigma Y^{\dagger}\X^{\dagger}
\]
and we wish to reduce the set of pre-images in $\X$ without loss.
We suppose $\Sigma$ is full rank.
\begin{enumerate}
\item Remove the pre-images for which a line of $Y\Sigma Y^{\dagger}$ is
null (and remove the corresponding column of $Y$). 
\item If the decomposition is not orthonormal, use a QR or LU decomposition
to find the null space of $\X^{\dagger}\X$, i.e. a full rank $Z$
such that $\X^{\dagger}\X Z=0$. We then remove $n$ pre-images (see
below) where $n$ is the rank of $Z$.
\item Finally, we remove non used pre-images like in step (1).
\end{enumerate}

\paragraph{Null space and pre-images}

We want to find $\X^{\prime}$ and $A$ such that $\X=\left(\begin{array}{cc}
\X^{\prime} & X^{\prime}A\end{array}\right)P$ where $P$ is a permutation matrix.

We have a basis $Z$ for the null subspace of $\X^{\dagger}\X$. If
$z$ is in the null subspace, then

\[
\forall i\,\X_{i}^{\dagger}\sum_{j}z_{j}\X_{j}=0\implies\sum_{j}z_{j}\X_{j}=0
\]
 since $\sum_{j}z_{j}\X_{j}$ belongs to the span of $\X$. 

To chose among the pre-images, we chose to remove first those that
are the less used, i.e. those for which $\left\Vert Y_{j\bullet}\right\Vert \left\Vert \X_{j}\right\Vert $
is minimum. We also have to ensure that $z_{j}$ is not too small,
i.e. is above $\delta\left\Vert z\right\Vert _{\infty}$. We then
remove entries one by one using the pivoted Gauss algorithm.

\subsubsection{Quadratic Programming approach (L1-optimisation)}

\label{sec:reduced-set}

Another to remove some pre-images is to try to directly optimise the
cost using an $L_{1}$ regulariser to set some rows of $Y$ close
to 0. Denoting $A=YZ$, we seek at minimising the difference 
\[
E=\left\Vert \X A\Sigma A^{\dagger}\X^{\dagger}-\X BTB^{\dagger}\X^{\dagger}\right\Vert ^{2}+\lambda\sum_{i}\left\Vert B_{i\bullet}\right\Vert _{\infty}
\]

Using $L_{1}$ regularisation ensures that $B$ is sparser than $A$
-- in particular, rows of $B$ are close to 0~(which means that the
corresponding pre-images $\X_{i}$ can be removed). 

In the following, we suppose that $A$ is of dimension $r\times n$
(i.e. $r$ basis vectors and $n$ feature vectors). Using the link
between the trace and the Frobenius norm, we have 
\[
E=\mbox{tr}\left(\trtimes{XA\Sigma A^{\dagger}X^{\dagger}-XBTB^{\dagger}X^{\dagger}}\right)
\]

Denoting $a_{i}=\X A_{\bullet i}\Sigma^{1/2}$ and $b_{i}=\X B_{\bullet i}T^{1/2}$
the two sets of vectors (in the feature space), we can then rewrite
$E$ as

\begin{align*}
E= & \mbox{tr}\left(\sum_{i,j}a_{i}a_{i}^{\dagger}a_{j}a_{j}^{\dagger}+b_{i}b_{i}^{\dagger}b_{j}b_{j}^{\dagger}-2a_{i}a_{i}^{\dagger}b_{j}b_{j}^{\dagger}\right)\\
= & \sum_{i,j}\left(a_{i}^{\dagger}a_{j}\right)^{2}+\left(b_{i}^{\dagger}b_{j}\right)^{2}-2\left|a_{i}^{\dagger}b_{j}\right|^{2}
\end{align*}

The problem we want to solve is linked to the ``reduced set'' approach
proposed in~\cite{Scholkopf1999Input}, where one seeks to minimise
the following cost function (with $L_{1}$ regularisation): 
\begin{align}
\mbox{minimise } & E_{RS}=\sum_{j}\nu_{j}\left\Vert a_{j}-b_{j}\right\Vert ^{2}+\lambda\xi_{i}\label{eq:rs-opt}\\
\mbox{subject to } & \forall i\,\xi_{i}\ge\max_{j}\left|B_{ij}\right|\nonumber 
\end{align}
The role of $\xi_{i}$ is to regularise the importances of feature
vectors; we need to set $\lambda$ appropriately so that some rows
of $B$ are close to 0 at the end of the optimisation. Finally, and
differently from other approaches, we added a new constant, $\nu_{i}$,
that ensures that $K\times E_{RS}\ge E$ for some $K\ge0$. We discuss
both in the following.

\paragraph{Relation with the reduced set approach}

We first check that minimising $E_{RS}$ solves our problem. The main
difference is that $E$ contains terms of the form $b_{i}^{\dagger}b_{j}$
and $a_{i}^{\dagger}b_{j}$. However, they will tend to be will be
close to 0 since the feature vectors will be approximately orthogonal.

Posing $b_{i}=\mu_{i}\left(a_{i}+c_{i}\right)$ with $a_{i}\bot c_{i}$,
we can first show that $\mu_{i}$ must be equal to $ $$\left\Vert a_{i}\right\Vert ^{2}\left(\left\Vert a_{i}\right\Vert ^{2}+\left\Vert c_{i}\right\Vert ^{2}\right)$
when $E_{RS}$ is minimised. Then, we can show that

\[
6\left\Vert a_{i}\right\Vert ^{2}\left\Vert a_{i}-b_{i}\right\Vert ^{2}\ge\left(a_{i}^{\dagger}a_{i}\right)^{2}+\left(b_{i}^{\dagger}b_{i}\right)^{2}-2\left|a_{i}^{\dagger}b_{i}\right|^{2}
\]

Now we have to prove that all cross terms ($i\not=j$) are minimised
if we minimise the new objective function, which intuitively is ensured
by the fact that $a_{i}\bot a_{j}$. Denoting $\Delta_{i}=a_{i}-b_{i}$,
we have for $i\not=j$:

\[
\left(a_{i}^{\dagger}a_{j}\right)^{2}+\left(b_{i}^{\dagger}b_{j}\right)^{2}-2\left|a_{i}^{\dagger}b_{j}\right|^{2}=\left(a_{i}^{\dagger}\Delta_{j}+\Delta_{i}^{\dagger}a_{j}+\Delta_{i}^{\dagger}\Delta_{j}\right)^{2}-2\left|a_{i}^{\dagger}\Delta_{j}\right|^{2}
\]
which is clearly bounded by $K\max_{i}\left\Vert \Delta_{i}\right\Vert ^{2}$
and hence by $K^{\prime}E_{RS}$.

Our problem is thus to optimise Eq. (\ref{eq:rs-opt}) with $\nu_{j}=\sigma_{j}=\left\Vert a_{j}\right\Vert ^{2}=|\Sigma_{jj}|$,
or equivalently, by posing $B_{\bullet j}^{\prime}=\sigma_{j}B_{j}$,
we can reformulate the optimisation problem as:

\begin{align}
\mbox{minimise } & E_{RS}=\sum_{j}\left\Vert \sigma_{j}a_{j}-b_{j}^{\prime}\right\Vert ^{2}+\lambda\xi_{j}\label{eq:rs-opt-1}\\
\mbox{subject to } & \forall i\,\xi_{i}\ge\max_{j}\sigma_{j}^{-1}\left|B_{ij}^{\prime}\right|\nonumber 
\end{align}

\paragraph{Setting $\lambda$}

If $B=A$ (trivial solution when $\lambda=0$), then to minimise the
above equation, we set $\xi_{i}=\max\sigma_{j}^{-1}|\sigma_{j}A_{ij}\sigma_{j}^{1/2}|$
and 
\[
E_{RS}^{(0)}=\lambda\sum_{i}\max_{j}|\sigma_{j}^{1/2}A_{ij}|
\]
If we remove the $i^{\nth}$ pre-image the error becomes

\[
E_{RS}^{(i)}=E_{RS}^{(0)}-\lambda\max_{j}\left|\text{\ensuremath{\sigma}}_{j}^{1/2}A_{ij}\right|+\sum_{j}\underbrace{\left\Vert \sigma_{j}^{3/2}\left(\X A_{\bullet j}-\X A_{\bullet j}^{(i)}\right)\right\Vert ^{2}}_{\sigma_{j}^{3}\left|K_{ii}A_{ij}\right|^{2}}
\]
where $A^{(i)}$ is $A$ with the $i^{\nth}$ row set to zero. Hence,
in order to remove the $i^{\nth}$ pre-image, we need to set $\lambda$
such that

\begin{align*}
E_{RS}^{(i)}-E_{RS}^{(0)} & =\left|K_{ii}\right|^{2}\sum_{j=1}^{r}\sigma_{j}^{3}A_{\bullet j}^{\dagger}KA_{\bullet j}\left|A_{ij}\right|^{2}-\lambda\max_{j}\left|\sigma_{j}^{1/2}A_{ij}\right|\ge0
\end{align*}
 where $K=k\left(\X,\X\right)$. If we want to remove (at least) $m$
pre-images whose indices are in $M$, we want to have

\[
\lambda\ge\frac{\sum_{i\in M}\left|K_{ii}\right|^{2}\sum_{j=1}^{r}\sigma_{j}^{3}A_{\bullet j}^{\dagger}KA_{\bullet j}\left|A_{ij}\right|^{2}}{\sum_{i\in M}\max_{j}\left|\sigma_{i}A_{ij}\right|}
\]
As an heuristic, we set $M$ to be the set of indices of pre-images
with ,minimum $E_{RS}^{(i)}-E_{RS}^{(0)}$.

\paragraph{Quadratic optimisation}

The quadratic programming problem can be solved using quadratic cone
optimisation. This is detailed in Appendix \ref{sec:qp-approach}.

\paragraph{Re-estimation of parameters}

We project the old operator into the new space in order to minimise
the error, i.e.

\[
\mathfrak{U}=\left(\Y BB^{\dagger}\Y^{\dagger}\X A\right)\Sigma A^{\dagger}\X^{\dagger}B\Y\Y^{\dagger}B^{\text{\ensuremath{\dagger}}}
\]

\section{Example}

\label{sec:example}

\begin{Verbatim}[commandchars=\\\{\},codes={\catcode`\$=3\catcode`\^=7\catcode`\_=8}]
\PY{c+cp}{\PYZsh{}}\PY{c+cp}{include <kqp}\PY{c+cp}{/}\PY{c+cp}{feature\PYZus{}matrix}\PY{c+cp}{/}\PY{c+cp}{dense.hpp>}
\PY{c+cp}{\PYZsh{}}\PY{c+cp}{include <kqp}\PY{c+cp}{/}\PY{c+cp}{kernel\PYZus{}evd}\PY{c+cp}{/}\PY{c+cp}{incremental.hpp>}
\PY{c+cp}{\PYZsh{}}\PY{c+cp}{include <kqp}\PY{c+cp}{/}\PY{c+cp}{probabilities.hpp>}

\PY{k+kt}{int} \PY{n}{main}\PY{p}{(}\PY{k+kt}{int}\PY{p}{,} \PY{k}{const} \PY{k+kt}{char}\PY{o}{*}\PY{o}{*}\PY{p}{)} \PY{p}{\PYZob{}}
    
    \PY{c+c1}{// --- Compute a density at random}
    
    \PY{c+c1}{// Definitions}
    \PY{k}{using} \PY{k}{namespace} \PY{n}{kqp}\PY{p}{;}
    \PY{k}{typedef} \PY{n}{Eigen}\PY{o}{:}\PY{o}{:}\PY{n}{Matrix}\PY{o}{<}\PY{k+kt}{double}\PY{p}{,} \PY{n}{Eigen}\PY{o}{:}\PY{o}{:}\PY{n}{Dynamic}\PY{p}{,} \PY{n}{Eigen}\PY{o}{:}\PY{o}{:}\PY{n}{Dynamic}\PY{o}{>} \PY{n}{Matrix}\PY{p}{;}
    \PY{k+kt}{int} \PY{n}{dim} \PY{o}{=} \PY{l+m+mi}{10}\PY{p}{;}
    
    \PY{c+c1}{// Creating an incremental builder}
    \PY{n}{IncrementalKernelEVD}\PY{o}{<}\PY{n}{DenseMatrix}\PY{o}{<}\PY{k+kt}{double}\PY{o}{>}\PY{o}{>} \PY{n}{kevd}\PY{p}{;}
    
    \PY{c+c1}{// Add 10 vectors with $\alpha_i=1$}
    \PY{k}{for}\PY{p}{(}\PY{k+kt}{int} \PY{n}{i} \PY{o}{=} \PY{l+m+mi}{0}\PY{p}{;} \PY{n}{i} \PY{o}{<} \PY{l+m+mi}{10}\PY{p}{;} \PY{n}{i}\PY{o}{+}\PY{o}{+}\PY{p}{)} \PY{p}{\PYZob{}}
        \PY{c+c1}{// Adds a random $\varphi_i$}
        \PY{n}{Matrix} \PY{n}{m} \PY{o}{=} \PY{n}{Matrix}\PY{o}{:}\PY{o}{:}\PY{n}{Random}\PY{p}{(}\PY{n}{dim}\PY{p}{,} \PY{l+m+mi}{1}\PY{p}{)}\PY{p}{;}
        \PY{n}{kevd}\PY{p}{.}\PY{n}{add}\PY{p}{(}\PY{n}{DenseMatrix}\PY{o}{<}\PY{k+kt}{double}\PY{o}{>}\PY{p}{(}\PY{n}{m}\PY{p}{)}\PY{p}{)}\PY{p}{;}
    \PY{p}{\PYZcb{}}
    
    \PY{c+c1}{// Get the result $\rho \approx X Y D Y^\dagger X^\dagger$}
    \PY{n}{DenseMatrix}\PY{o}{<}\PY{k+kt}{double}\PY{o}{>} \PY{n}{mX}\PY{p}{;}
    \PY{k}{typename} \PY{n}{AltDense}\PY{o}{<}\PY{k+kt}{double}\PY{o}{>}\PY{o}{:}\PY{o}{:}\PY{n}{type} \PY{n}{mY}\PY{p}{;}
    \PY{n}{Eigen}\PY{o}{:}\PY{o}{:}\PY{n}{Matrix}\PY{o}{<}\PY{k+kt}{double}\PY{p}{,} \PY{n}{Eigen}\PY{o}{:}\PY{o}{:}\PY{n}{Dynamic}\PY{p}{,}\PY{l+m+mi}{1}\PY{o}{>}  \PY{n}{mD}\PY{p}{;}
    
    \PY{n}{kevd}\PY{p}{.}\PY{n}{get\PYZus{}decomposition}\PY{p}{(}\PY{n}{mX}\PY{p}{,} \PY{n}{mY}\PY{p}{,} \PY{n}{mD}\PY{p}{)}\PY{p}{;}

    \PY{c+c1}{// --- Compute a kEVD for a subspace}
    
    \PY{n}{IncrementalKernelEVD}\PY{o}{<}\PY{n}{DenseMatrix}\PY{o}{<}\PY{k+kt}{double}\PY{o}{>}\PY{o}{>} \PY{n}{kevd\PYZus{}event}\PY{p}{;}
    \PY{k}{for}\PY{p}{(}\PY{k+kt}{int} \PY{n}{i} \PY{o}{=} \PY{l+m+mi}{0}\PY{p}{;} \PY{n}{i} \PY{o}{<} \PY{l+m+mi}{3}\PY{p}{;} \PY{n}{i}\PY{o}{+}\PY{o}{+}\PY{p}{)} \PY{p}{\PYZob{}}
        \PY{c+c1}{// Adds a random $\varphi_i$}
        \PY{n}{Matrix} \PY{n}{m} \PY{o}{=} \PY{n}{Matrix}\PY{o}{:}\PY{o}{:}\PY{n}{Random}\PY{p}{(}\PY{n}{dim}\PY{p}{,} \PY{l+m+mi}{1}\PY{p}{)}\PY{p}{;}
        \PY{n}{kevd\PYZus{}event}\PY{p}{.}\PY{n}{add}\PY{p}{(}\PY{n}{DenseMatrix}\PY{o}{<}\PY{k+kt}{double}\PY{o}{>}\PY{p}{(}\PY{n}{m}\PY{p}{)}\PY{p}{)}\PY{p}{;}
    \PY{p}{\PYZcb{}}

    
    \PY{c+c1}{// --- Compute some probabilities}
    
    \PY{c+c1}{// Setup densities and events}
    \PY{n}{Density}\PY{o}{<}\PY{n}{DenseMatrix}\PY{o}{<}\PY{k+kt}{double}\PY{o}{>}\PY{o}{>} \PY{n}{rho}\PY{p}{(}\PY{n}{kevd}\PY{p}{)}\PY{p}{;}
    \PY{n}{Event}\PY{o}{<}\PY{n}{DenseMatrix}\PY{o}{<}\PY{k+kt}{double}\PY{o}{>}\PY{o}{>} \PY{n}{event}\PY{p}{(}\PY{n}{kevd\PYZus{}event}\PY{p}{)}\PY{p}{;}
    
    \PY{c+c1}{// Compute the probability}
    \PY{n}{std}\PY{o}{:}\PY{o}{:}\PY{n}{cout} \PY{o}{<}\PY{o}{<} \PY{l+s}{"}\PY{l+s}{Probability = }\PY{l+s}{"} \PY{o}{<}\PY{o}{<} \PY{n}{rho}\PY{p}{.}\PY{n}{probability}\PY{p}{(}\PY{n}{event}\PY{p}{)} \PY{o}{<}\PY{o}{<} \PY{n}{std}\PY{o}{:}\PY{o}{:}\PY{n}{endl}\PY{p}{;}

    \PY{c+c1}{// Conditional probability}
    \PY{n}{Density}\PY{o}{<}\PY{n}{DenseMatrix}\PY{o}{<}\PY{k+kt}{double}\PY{o}{>}\PY{o}{>} \PY{n}{rho\PYZus{}cond} \PY{o}{=} \PY{n}{event}\PY{p}{.}\PY{n}{project}\PY{p}{(}\PY{n}{kevd}\PY{p}{)}\PY{p}{.}\PY{n}{normalize}\PY{p}{(}\PY{p}{)}\PY{p}{;} 

    \PY{c+c1}{// Conditional probability (orthogonal event)}
    \PY{n}{Density}\PY{o}{<}\PY{n}{DenseMatrix}\PY{o}{<}\PY{k+kt}{double}\PY{o}{>}\PY{o}{>} \PY{n}{rho\PYZus{}cond\PYZus{}orth} \PY{o}{=} \PY{n}{event}\PY{p}{.}\PY{n}{project}\PY{p}{(}\PY{n}{kevd}\PY{p}{,} \PY{k+kc}{true}\PY{p}{)}\PY{p}{.}\PY{n}{normalize}\PY{p}{(}\PY{p}{)}\PY{p}{;}
    
    \PY{k}{return} \PY{l+m+mi}{0}\PY{p}{;}
\PY{p}{\PYZcb{}}
\end{Verbatim}

\section{Conclusion}

This document described the Kernel Quantum Probability Library, that
can be used to compute quantum events and density in the an arbitrary
feature space and relies only on the definition of a kernel, i.e.
of the inner product between any two feature vectors.

\bibliographystyle{abbrv}
\bibliography{kqp}

\appendix

\section{QP Approach}

\label{sec:qp-approach}

In this section, we derive a computationally efficient way to optimise
Eq. (\ref{eq:rs-opt}). We show here how to transform this optimisation
problem into a cone quadratic programming approach proposed in \cite{Andersen:2011wx}.
We handle both the complex and the real field cases.

\subsection{Precomputations}

Writing $\alpha_{i}=\sigma_{i}A_{\bullet i}$, $\beta_{i}=\tau_{i}B_{\bullet i}$
and $K=\X^{\dagger}\X$ the gram matrix ($r$ is the rank of the operator,
$n$ is the number of pre-images), we have

\begin{align*}
E_{RS}= & \sum_{q=1}^{r}\beta_{q}^{\dagger}\beta_{q}-2\nu_{q}\Re\left(\alpha_{q}^{\dagger}\beta_{q}\right)+\lambda\xi_{q}\\
= & \sum_{q=1}^{r}\beta_{q}^{\dagger}K\beta_{q}-2\Re\left(\alpha_{q}^{\dagger}K\beta_{q}\right)+\lambda\xi_{q}
\end{align*}

We get back to a real case by posing $\beta_{q}=\beta_{q}^{\prime}+i\beta_{q}^{\prime\prime}$
and $\alpha_{q}=\alpha_{q}^{\prime}+i\alpha_{q}^{\prime\prime}$.
Dropping $q$ for clarity, we have

\begin{align*}
\beta^{\dagger}K\beta & =\beta^{\prime\dagger}K\beta^{\prime}+\beta^{\second\dagger}K\beta^{\second}+i\left(\beta^{\prime\dagger}K\beta^{\second}-\beta^{\second\dagger}K\beta^{\prime}\right)\\
 & =\beta^{\prime\dagger}\Re\left(K\right)\beta^{\prime}+\beta^{\second\dagger}\Re\left(K\right)\beta^{\second}+i\left(\beta^{\prime\dagger}K\beta^{\second}-\beta^{\prime\dagger}K^{\dagger}\beta^{\second}\right)\\
 & =\beta^{\prime\dagger}\Re\left(K\right)\beta^{\prime}+\beta^{\second\dagger}\Re\left(K\right)\beta^{\second}-2\beta^{\prime\dagger}\Im\left(K\right)\beta^{\second}
\end{align*}

and

\begin{align*}
\Re\left(\alpha^{\dagger}K\beta\right) & =\Re\left(\alpha^{\prime\dagger}K\beta^{\prime}+\alpha^{\second\dagger}K\beta^{\second}+i\alpha^{\prime\dagger}K\beta^{\second}-i\alpha^{\second\dagger}K\beta^{\prime}\right)\\
 & =\left(\alpha^{\prime\dagger}\Re\left(K\right)+\alpha^{\second\dagger}\Im\left(K\right)\right)\beta^{\prime}+\left(\alpha^{\second\dagger}\Re\left(K\right)-\alpha^{\prime\dagger}\Im\left(K\right)\right)\beta^{\second}
\end{align*}

Hence
\[
\beta_{q}^{\dagger}K\beta_{q}-2\Re\left(\alpha_{q}^{\dagger}K\beta_{q}\right)+\lambda\xi_{q}=\left(\begin{array}{c}
\beta_{q}^{\prime}\\
\beta_{q}^{\second}
\end{array}\right)^{\dagger}\left(\begin{array}{cc}
\Re\left(K\right) & \Im(K)\\
\Im(K) & \Re(K)
\end{array}\right)\left(\begin{array}{c}
\beta_{q}^{\prime}\\
\beta_{q}^{\second}
\end{array}\right)+\left(\begin{array}{c}
\alpha^{\prime}\\
\alpha^{\second}
\end{array}\right)^{\dagger}\left(\begin{array}{cc}
\Re(K) & \Im(K)\\
-\Im(K) & \Re(K)
\end{array}\right)\left(\begin{array}{c}
\beta_{q}^{\prime}\\
\beta_{q}^{\second}
\end{array}\right)
\]

If we let
\begin{eqnarray*}
x & = & \left(\begin{array}{cccccc}
\beta_{1}^{\dagger} & \cdots & \beta_{r}^{\dagger} & \xi_{1} & \cdots & \xi_{n}\end{array}\right)^{\dagger}
\end{eqnarray*}
with $\beta_{i}=\left(\begin{array}{c}
\beta_{i}^{\prime}\\
\beta_{i}^{\second}
\end{array}\right)$ in the complex case and $\beta_{i}=\beta_{i}^{\prime}$ in the real
one.

We require that both the real and imaginary part be inferior to $\xi$,
i.e. that

\begin{align*}
\forall i\in1\dots n,\,\forall q\in1\dots r,\, & \nu_{q}\left(\pm\beta_{qi}^{\prime}\pm\beta_{qi}^{\second}\right)+\xi_{i}\ge0
\end{align*}
where $\nu_{q}$ are weights associated to basis vectors in the feature
space, and $x$ has a length $n\times(r^{\prime}+1)$. Our problem
can be expressed as a cone quadratic problem

\begin{align*}
\mbox{minimise } & x^{\dagger}Hx+2c^{\dagger}x\\
\mbox{subject to } & Gx\le0
\end{align*}
Denoting $\Id_{n}^{(.)}$ the matrix $(\begin{array}{ccc}
\Id_{n} & \cdots & \Id_{n}\end{array})^{\dagger}$ and $\mathbf{1}^{(.)}$ the matrix $(\begin{array}{ccc}
1 & \cdots & 1\end{array})^{\dagger}$, we can identify: 
\begin{eqnarray*}
H & = & \left(\begin{array}{cc}
K_{\times r}^{\prime}\\
 & \mathbf{0}_{n}
\end{array}\right)\\
c & = & \left(\begin{array}{c}
-K^{\second}\alpha_{1}\\
\vdots\\
-K^{\second}\alpha_{r}\\
\frac{\lambda}{2}\mathbf{1}^{(n)}
\end{array}\right)\\
G & = & \left(\begin{array}{cc}
-S & -\Id_{n}^{(r^{\prime})}\\
S & -\Id_{n}^{(r^{\prime})}
\end{array}\right)
\end{eqnarray*}
where $\mbox{diag}_{r}$ repeats the matrix $r$ times in the diagonal
where 
\[
S=\mbox{diag}\left(\nu_{1}G_{0},\ldots\nu_{1}G_{0}\right)
\]
with $G_{0}$ defined latter.

\paragraph{Case $\mathbb{K}=\mathbb{R}$}

In the case where $\mathbb{K}=\mathbb{R}$, we have $K^{\prime}=K^{\second}=K$
and $r^{\prime}=r$ and $G_{0}=\Id_{n}$

\paragraph{Case $\mathbb{K}=C$}

we have $r^{\prime}=2r$ and

\begin{align*}
K^{\prime} & =\left(\begin{array}{cc}
\Re\left(K\right) & -\Im\left(K\right)\\
-\Im\left(K\right) & \Re\left(K\right)
\end{array}\right)\\
K^{\second} & =\left(\begin{array}{cc}
\Re\left(K\right) & \Im\left(K\right)\\
-\Im\left(K\right) & \Re\left(K\right)
\end{array}\right)\\
G_{0} & =\left(\begin{array}{cc}
\Id_{n} & \Id_{n}\\
\Id_{n} & -\Id_{n}
\end{array}\right)
\end{align*}

\subsection{Pre-solving the system}

In order to speed up, we need to solve the linear systems defined
by

\[
\left(\begin{array}{cc}
H & G^{\dagger}\\
G & V
\end{array}\right)
\]
where $V$ is a diagonal negative matrix. With a bit of re-ordering,
this gives
\[
\left(\begin{array}{cccc}
\mbox{diag}_{r}K^{\prime} & -\Id_{nr^{\prime}} & \Id_{n}^{(r^{\prime})} & \mathbf{0}_{n}^{(r^{\prime})}\\
-S & -U & \mathbf{0}_{nr^{\prime}} & -\Id_{n}^{(r^{\prime})}\\
S & \mathbf{0}_{2nr} & -V & -\Id_{n}^{(r^{\prime})}\\
\mathbf{0}_{n}^{(r^{\prime})\dagger} & -\Id_{n}^{(r^{\prime})\dagger} & -\Id_{n}^{(2r)\dagger} & \mathbf{0}
\end{array}\right)\left(\begin{array}{c}
x\\
z\\
t\\
y
\end{array}\right)=\left(\begin{array}{c}
a\\
b\\
d\\
c
\end{array}\right)
\]
where $U$ and $V$ are positive semi-definite (diagonal) matrices
of size $2nr$.

We want to perform a $LDL^{\dagger}$ decomposition of this matrix
(a D-Cholesky). Given the structure of the above matrix, we decompose
these matrices as

\[
L=\left(\begin{array}{cccc}
L_{11}\\
L_{21} & L_{22}\\
L_{31} & L_{32} & L_{33}\\
L_{41} & L_{42} & L_{43} & L_{44}
\end{array}\right)\mbox{ and }D=\left(\begin{array}{cccc}
D_{1}\\
 & D_{2}\\
 &  & D_{3}\\
 &  &  & D_{4}
\end{array}\right)
\]

\paragraph{Solving $L_{11}$}

A Cholesky decomposition of $K^{\prime}$, $AA^{\dagger}=K^{\prime}$
gives

\[
L_{11}=\mbox{diag}\left(\begin{array}{ccc}
A & \cdots & A\end{array}\right)\mbox{ and }D_{1}=\Id
\]

Note that in the complex field case, we can decompose the problem
into
\begin{align*}
A_{11}A_{11}^{\dagger} & =\Re\left(K\right)\\
A_{21}A_{11}^{\dagger} & =-\Im\left(K\right)\\
A_{22}A_{22}^{\dagger} & =\Re\left(K\right)-A_{21}A_{21}^{\dagger}
\end{align*}

\paragraph{Solving $L_{21}$ and $L_{31}$}

We now have 
\[
\left(\begin{array}{c}
L_{21}\\
L_{31}
\end{array}\right)L_{11}^{\dagger}=\left(\begin{array}{c}
-S\\
S
\end{array}\right)
\]
where $S=\mbox{diag}\left(\nu_{1}G_{0},\ldots\nu_{1}G_{0}\right)$.
This can be solved straightforwardly by first solving%
\footnote{Note that $BB^{\dagger}$ is positive definite since $BB^{\dagger}=A^{-\dagger}G_{0}^{2}A^{-1}$
where $G_{0}$ is positive definite%
} $BA^{\dagger}=G_{0}$ .

\[
L_{21}=\mbox{diag}\left(\begin{array}{ccc}
-\nu_{1}B & \cdots & -\nu_{r}B\end{array}\right)\mbox{ and }L_{31}=\mbox{diag}\left(\begin{array}{ccc}
\nu_{1}B & \cdots & \nu_{r}B\end{array}\right)
\]

\paragraph{Solving $L_{22}$}

We have to solve $L_{22}D_{22}^{\dagger}L_{22}=-U-L_{21}L_{21}^{\dagger}=-\mbox{diag}\left(\left(U_{i}+\nu_{i}^{2}BB^{\dagger}\right)_{i}\right)$.
Since $U_{i}+\nu_{i}^{2}BB^{\dagger}$ is positive definite, it is
sufficient to solve the $r$ Cholesky decompositions $L_{22}^{(i)}L_{22}^{(i)\dagger}=U_{i}+\nu_{i}^{2}BB^{\dagger}$
with

\[
L_{22}=\mbox{diag}\left(\begin{array}{ccc}
L_{22}^{(1)} & \cdots & L_{22}^{(r)}\end{array}\right)\mbox{ and }D_{2}=-\Id_{nr^{\prime}}
\]

\paragraph{Solving $L_{32}$}

Then, we find $L_{32}$ by solving $L_{32}D_{2}L_{22}^{\dagger}=-L_{31}L_{21}^{\dagger}$
which can be solved by solving the $r$ systems $L_{32}^{(i)}L_{22}^{(i)^{\dagger}}=-\nu_{i}^{2}BB^{\dagger}$, 

\[
L_{32}=\mbox{diag}\left(\begin{array}{ccc}
L_{32}^{(1)} & \cdots & L_{32}^{(r)}\end{array}\right)
\]

\paragraph{Solving $L_{33}$}

For $L_{33}$, we have 

\[
L_{33}D_{3}L_{33}^{\dagger}-L_{32}L_{32}^{\dagger}+L_{31}L_{31}^{\dagger}=-V
\]
 which can be solved by $r$ D-Cholesky decomposition (all the matrices
are block diagonal):

\begin{equation}
L_{33}^{(i)}D_{3}^{(i)}L_{33}^{(i)\dagger}=-V_{i}-\nu_{i}^{2}BB^{\dagger}+L_{32}^{(i)}L_{32}^{(i)\dagger}\label{eq:L33}
\end{equation}

In order to simplify the computation, let us prove that the right
hand side is negative definite; we have

\begin{align*}
L_{32}^{(i)}L_{32}^{(i)\dagger} & =-\nu_{i}^{2}BB^{\dagger}L_{22}^{(i)-1}L_{22}^{(i)-\dagger}BB^{\dagger}\\
 & =-\nu_{i}^{2}BB^{\dagger}\left(U_{i}+BB^{\dagger}\right)^{-1}BB^{\dagger}\\
 & =\nu_{i}^{2}BB^{\dagger}-U_{i}+U_{i}\left(U_{i}+BB^{\dagger}\right)^{-1}U_{i}
\end{align*}

We have

\[
U_{i}^{-1}\left(U_{i}+\nu_{i}^{2}BB^{\dagger}\right)U_{i}^{-1}=U_{i}^{-1}+\nu_{i}^{2}U_{i}^{-1}BB^{\dagger}U_{i}^{-1}\ge U_{i}^{-1}>0
\]
 using the partial order of definite matrices and its properties.
This implies that

\[
U_{i}\ge U_{i}\left(U_{i}+\nu_{i}^{2}BB^{\dagger}\right)^{-1}U_{i}>0
\]
 and hence $-U_{i}+U_{i}\left(U_{i}+\nu_{i}^{2}BB^{\dagger}\right)^{-1}U_{i}$
is positive semi-definite, which in turn implies that the right hand
side of Eq. (\ref{eq:L33}) is negative definite. This shows that
we can find $L_{33}$ using a Cholesky decomposition

\[
L_{33}^{(i)}L_{33}^{(i)\dagger}=V_{i}+\nu_{i}^{2}BB^{\dagger}-L_{32}^{(i)}L_{32}^{(i)\dagger}
\]
and that $D_{3}=\Id_{rn}$.

\paragraph{Solving $L_{41}$, $L_{42}$ and $L_{43}$}

For the fourth row, we first have trivially $L_{41}=0$.

To solve $L_{42}D_{2}L_{22}^{\dagger}=\left(\begin{array}{ccc}
-\Id_{n} & \cdots & -\Id_{n}\end{array}\right)$, we have to solve $L_{42}^{(i)}L_{22}^{(i)\dagger}=\Id_{n}$ for
$i=1\dots r$, and then

\[
L_{42}=\left(\begin{array}{ccc}
L_{42}^{(1)} & \cdots & L_{42}^{(r)}\end{array}\right)
\]

To find $L_{43}$, we have to solve $L_{43}^{(i)}L_{33}^{(i)\dagger}=\Id_{n}-L_{42}^{(i)}L_{32}^{(i)\dagger}$
for $i=1\dots r$ and

\[
L_{43}=\left(\begin{array}{ccc}
L_{43}^{(1)} & \cdots & L_{43}^{(r)}\end{array}\right)
\]

Finally, the last equation $L_{42}D_{2}L_{42}^{\dagger}+L_{43}D_{3}L_{43}^{\dagger}+L_{44}D_{4}L_{44}^{\dagger}=0$
can be solved by computing the Cholesky decomposition

\[
L_{44}L_{44}^{\dagger}=\sum_{i=1}^{r}L_{42}^{(i)}L_{42}^{(i)\dagger}+L_{43}^{(i)}L_{43}^{(i)\dagger}\mbox{ with }D_{4}=\Id_{n}
\]

\subsection{Solving the linear system}

Here is the final structure of the decomposition:
\begin{align*}
L & =\left(\begin{array}{cccc}
A_{\times r}\\
-L_{21} & L_{22}\\
L_{21} & L_{32} & L_{33}\\
\mathbf{0} & L_{42} & L_{43} & L_{44}
\end{array}\right)\\
D & =\left(\begin{array}{ccc}
\Id_{nr}\\
 & -\Id_{2rn}\\
 &  & \Id_{n}
\end{array}\right)
\end{align*}
with 
\[
L_{21}=\mbox{diag}\left(\begin{array}{ccc}
-\nu_{1}B & \cdots & -\nu_{r}B\end{array}\right)
\]
\[
\]

which gives the following systems to solve:

\begin{align*}
A^{\text{}}x_{i}^{\prime} & =a_{i}\\
L_{22}^{(i)}z_{i}^{\prime} & =-\nu_{i}Bx_{i}^{\prime}-c_{i}\\
L_{33}^{(i)}t_{i}^{\prime} & =\nu_{i}Bx_{i}^{\prime}-L_{32}^{(i)}z_{i}^{\prime}-d_{i}\\
L_{44}y^{\prime} & =b+\sum_{i=1}^{r}L_{43}^{(i)}t_{i}^{\prime}+L_{42}^{(i)}z_{i}^{\prime}
\end{align*}

and finally
\begin{align*}
L_{44}^{\dagger}y & =y^{\prime}\\
L_{33}^{(i)\dagger}t_{i} & =t_{i}^{\prime}-L_{43}^{(i)\dagger}y\\
L_{22}^{(i)\dagger}y_{i} & =y_{i}^{\prime}-L_{32}^{(i)\dagger}t_{i}-L_{42}^{(i)\dagger}y\\
A^{\dagger}x_{i} & =x_{i}^{\prime}-\nu_{i}B^{\dagger}t_{i}+\nu_{i}B^{\dagger}z_{i}
\end{align*}

\end{document}